\documentclass[12pt]{article}
\usepackage{latexsym}
\usepackage{amsmath}
\usepackage{wasysym}
\usepackage{amsfonts}
\usepackage{dsfont}
\usepackage{pifont}
\usepackage{bm}
\usepackage{epsf}
\usepackage{epsfig,graphicx}
\usepackage{amsmath,amssymb,bm,epsf,epsfig,graphicx}

\usepackage{color}
\definecolor{rougef}{rgb}{0.56,0,0}
\definecolor{vertf}{rgb}{0,0.5,0}
\definecolor{bleuf}{rgb}{0,0,0.8}
\definecolor{violetf}{rgb}{0.5,0,0.5}

\usepackage[vcentermath]{youngtab}

\usepackage[colorlinks=true,linkcolor=blue,citecolor=blue,urlcolor=black,filecolor=black,linktocpage=true]{hyperref}

 \csname
@addtoreset\endcsname{equation}{section}

\def\pe{\prime}
\def\3s{{s \choose 3}}
\def\4s{{s \choose 4}}
\def\5s{{s \choose 5}}
\def\6s{{s \choose 6}}

\def\12{\frac{1}{2}}
\def\fr{\frac}
\def\ft{\footnote}

\def\1{\ell_1}
\def\2{\ell_2}
\def\ss{\ell_s}

\def\pr{\partial}

\def\prd{\partial \cdot}

\def\be{\begin{equation}}
\def\ee{\end{equation}}
\def\bea{\begin{eqnarray}}
\def\eea{\end{eqnarray}}
\def\ba{\begin{array}}
\def\ea{\end{array}}
\def\bec{\begin{center}}
\def\ec{\end{center}}

\def\g{\gamma} 

\def\d{\delta}

\def\Th{\Theta}

\def\l{\lambda}
\def\L{\Lambda}
\def\m{\mu}
\def\n{\nu}

\def\vf{\varphi}

\def\tvdots{\smash\vdots}

\def\cF{{\cal F}}

\def\cH{{\cal H}}

\def\cO{{\cal O}}

\def\cR{{\cal R}}

\begin{document}

\begin{center}


{\large\sc \bf Mixed-symmetry multiplets and higher-spin curvatures}\\


\vspace{30pt} {\sc Xavier Bekaert${}^{\; a}$, Nicolas Boulanger${}^{\; b}$ and Dario Francia${}^{\; c}$} \\ \vspace{10pt} 
{${}^a$\small {\sl Laboratoire de Math\'ematiques et Physique Th\'eorique \\
Unit\'e Mixte de Recherche 7350 du CNRS \\
F\'ed\'eration de Recherche 2964 Denis Poisson \\
Universit\'e Fran{\c{c}}ois Rabelais, Parc de Grandmont \\
F-37200 Tours, France} \\
{\texttt{Xavier.Bekaert@lmpt.univ-tours.fr}}\\
${}^b$ {\sl Service de M\'ecanique et Gravitation \\
Universit\'e de Mons, UMONS \\
20 Place du Parc \\
B-7000 Mons, Belgium}\\
{\texttt{nicolas.boulanger@umons.ac.be}}\\
${}^c${\sl Scuola Normale Superiore and INFN \\ Piazza dei Cavalieri 7, I-56126 Pisa, Italy} }\\
{\texttt{dario.francia@sns.it}}\\
\vspace{10pt}


\vspace{30pt} {\sc\large Abstract}\end{center}
We study the higher-derivative equations for gauge potentials of arbitrary mixed-symmetry type obtained by  setting to zero the divergences of the corresponding curvature tensors. We show that they propagate the same reducible multiplets as the Maxwell-like second-order equations for gauge fields subject to constrained gauge transformations. As an additional output of our analysis, we provide a streamlined presentation of the Ricci-like case, where the traces of the same curvature tensors are set to zero, and we present a simple algebraic evaluation of the particle content associated with the Labastida and with the Maxwell-like second-order equations. 

\vfill
\setcounter{page}{1}

\pagebreak




\section{Introduction}\label{sec:intro}


  Higher-spin theories call for higher derivatives. While this is known to be an intrinsic feature of the interactions among such systems, as a matter of fact the same peculiarity also manifests itself for free massless particles of spin higher than $2$, due to the structure of the corresponding curvature tensors. Free equations of motion can indeed be formulated exploiting customary, second-order  differential operators, but only at the price of constraining some of the components of the  gauge parameters to vanish.  As we shall stress, this feature can be interpreted as arising from the partial gauge fixing of equations originating directly from fully gauge invariant curvatures.
  
 For an arbitrary $GL(D)-$reducible gauge potential $\vf$  subject to gauge transformations of the form\ft{Gauge parameters are denoted with $\L_{\, i}$, where the lower index $i$ stands for one missing space-time index in the $i$th family. Correspondingly, the operators $d^{\, i}$ with upper index $i$ denote exterior derivatives acting on the $i$th family. The Einstein convention is enforced so that  \eqref{gauge} defines a  scalar quantity in family-index space. $T_{\, ij}$ and $d_{\, i}$ denote operators computing traces and divergences, respectively, in the corresponding families $i, j = 1,  \ldots, s$. See also Appendix \ref{A1}.}
\be \label{gauge}
\d \, \vf \, = \, d^{\, i} \, \L_{\, i} \, ,
\ee
with reducible gauge parameters $\L_{\, i}$, two types of wave equations can be assigned.  The Labastida equations \cite{labastida}, together with the corresponding trace conditions on the gauge parameters, generalise Fronsdal's equations \cite{fronsdal} for symmetric tensors and are usually regarded as providing the standard covariant description of the massless particles (i.e. the irreps of the little group of the light-like momentum, $O (D-2)$) formally associated with $\vf$ : 
\be \label{laba}
\begin{split}
& \cF \, := \, \Box \vf \, - \, d^{\, i} \, d_{\, i} \, \vf \, + \, \frac{1}{2} \, d^{\, i}\, d^{\, j} \, T_{\, i j} \, \vf \, = \, 0 \, , \\
& T_{\, (i j} \, \L_{\, k)} \, = \, 0\, .
\end{split}
\ee
In particular, for tensors $\vf_{_Y}$ taking values in {\it irreducible} representations of $GL(D)$, equations \eqref{laba} provide a description of {\it single-particle}, massless degrees of freedom. In the latter case the gauge transformation \eqref{gauge} should be properly projected.

More recently, second-order Maxwell-like equations were proposed for arbitrary tensors of $GL(D)$ in \cite{M}, with gauge parameters subject to suitable differential conditions providing somehow the counterparts of the Labastida trace constraints\ft{Let us also mention, however, that presence or absence of constraints also depends on how one chooses to parametrise the gauge symmetry of a given differential operator. For the symmetric sector of \eqref{maxwell}, for instance, it was shown in \cite{FLS} how to describe the same amount of gauge invariance by means of fully unconstrained, higher-derivative and reducible gauge transformations.}:
\be \label{maxwell}
\begin{split}
& M \, := \, \Box \vf \, - \, d^{\, i} \, d_{\, i} \, \vf \,  = \, 0 \, , \\
& d^{\, i} \, d^{\, j} d_{\, (i} \, \L_{\, j)} \, = \, 0\, .
\end{split}
\ee
Differently from \eqref{laba}, these equations propagate all the massless particles contained in the associated reducible tensor of $GL(D-2)$, i.e. all the $O(D-2)-$components contained in the corresponding branching. Indeed, even if enforced on irreducible $GL(D)-$tensors $\vf_{_Y}$, the Maxwell-like equations \eqref{maxwell} still describe a reducible spectrum, corresponding to the multiplet of particles contained in the $GL(D-2)-$tableau formally corresponding to $\vf_{_Y}$. Thus, for instance, for the case of symmetric tensors, together with a massless particle of spin $s$  the equations \eqref{maxwell} also propagate additional massless particles of spin $s-2$, $s-4$, and so on, down to $s=1$ or to $s=0$ depending on the parity of $s\,$.

All the results of our paper equally apply to both reducible and irreducible tensors of $GL(D)$. On general grounds, however, the choice of working with reducible tensors, while also  leading to some formal simplifications (like dispensing with the need to perform projections), is actually more natural from the string-theoretical vantage point where physical fields emerge as coefficients of products of bosonic oscillators.  Reducible higher-spin systems have been less extensively studied in the literature.  Still we regard them as being worth of interest for a number of reasons. First,  they provide an alternative option equally viable in principle with respect to the standard irreducible  higher-spin models. In addition, one observes that discarding the requirement of irreducibility typically  leads to various technical simplifications. This is especially true for the Maxwell-like models  \eqref{maxwell}, essentially due to the fact that one does not need to explicitly deal with traces.  Finally, the same spectrum described by \eqref{maxwell} emerges when considering free tensionless  strings \cite{triplets1, triplets3, triplets2, fs2, st}, to which equations \eqref{maxwell} are indeed directly related, as discussed in \cite{M}.  This connection in particular provides a clear physical motivation for studying theories related to  \eqref{maxwell}, with the perspective that proceeding along this path one could shed some light on  the still rather mysterious relation between interacting massless higher spins and strings. For other approaches to reducible higher-spin models see e. g. \cite{FT, SV, FST}.

In the present paper we shall explore  how to recover equations \eqref{laba} and \eqref{maxwell} starting from suitable curvatures, defined as tensors that are identically gauge-invariant under \eqref{gauge} and that do not vanish when equations \eqref{laba} or \eqref{maxwell} hold. Such higher-spin curvatures were first introduced in \cite{dwf}  for symmetric tensors\ft{See also the earlier work \cite{weinberg} for on-shell $SO(3,1)-$analogues of the off-shell $GL(D)$ curvatures of \cite{dwf}.}, and later extended in \cite{Hull:2001iu,bb1,de Medeiros:2002ge} to the general class of gauge potentials with mixed symmetry. In particular, if $\vf_{_Y}$ is a $GL(D)-$irreducible field associated with a Young tableau $Y$ possessing $s$ columns --- in the basis with manifest antisymmetry among indices placed along columns, whose lengths we shall denote with $\ell_1$, $\ell_2$, \ldots, $\ell_s$ --- its curvature $\cR \, (\vf_{_Y})$ is again associated with a  Young tableau in $GL(D)$,  with one additional upper row corresponding to the $s$ curls entering its definition, 
\be
\cR\,(\vf_{_Y}) := d^{1} \ldots d^{s} \,\vf_{_Y}\,.
\ee
Similarly, for reducible {\it multi-form} fields $\vf$, taking values in tensor products of forms of various degrees, curvatures $\cR \, (\vf)$ can be defined analogously as multi-forms with the same number of factors, each corresponding to a form of degree augmented by one \cite{henneauxdb}. Thus, if one takes the number of columns of a tableau as a possible generalisation of the concept of spin, we see that for all gauge potentials of spin higher than $2$ their curvatures contain more than two derivatives.

The equations obtained setting to zero the traces of the higher-spin curvatures,
\be \label{traceR}
T_{\, i j } \, \cR \, (\vf) \, = \, 0 \, \hskip 2cm i, \, j \, = \, 1, \, \ldots, \, s
\ee
can be regarded as backbone equations for gauge theories of any spin, generalising the condition of vanishing Ricci tensor for the massless spin-$2$ particle. Indeed, in spite of the different number of derivatives involved, it is always possible to relate \eqref{traceR} to the Fronsdal-Labastida second-order equations \eqref{laba}. This was first proven for the case of irreducible tensors $\vf_{_Y}$ in \cite{bb2,Bekaert:2006ix}, while we shall show in the present work how to extend the equivalence to cover the case of reducible tensors. 

Moreover, even part of Vasiliev's system of non-linear differential equations for totally-symmetric gauge fields \cite{Vasiliev:1990en} can be interpreted as a consistent deformation of the frame-like counterpart of \eqref{traceR}, there written as the on-shell equality between the higher-spin curvatures and their traceless, or ``Weyl'', component. 

In the class of bosonic fields the only two representations that are not covered by \eqref{traceR} are $p-$forms. In particular, scalar fields, being gauge invariant, can be considered as representing their own curvature tensors, in a sense, on which one can impose the mass-shell condition $p^{\, 2} = 0$. On the other hand, the basic equation for massless $p-$forms involves {\it the divergence} of the corresponding field strength. While it is true that once \eqref{traceR} is imposed all divergences of $\cR \, (\vf)$  are also forced to vanish, due to the Bianchi identities that curvatures satisfy, one might also wonder whether it makes sense to impose on higher-spin curvatures the condition of vanishing divergence in itself, not in conjunction with the Ricci-like equation \eqref{traceR}.

Thus, the main goal of this paper is to study the transversality conditions
\be \label{divR}
d_{\, i} \, \cR \, (\vf) \, = \, 0  \, \hskip 2cm i,  \, = \, 1, \, \ldots, \, s
\ee
together with their counterparts for irreducible tensors, $d_{\, i} \, \cR \, (\vf_{_Y}) = 0$, and to show their equivalence to the second-order equations \eqref{maxwell}.  For the class of symmetric tensors this task was performed in \cite{divR}, where it was shown that the corresponding higher-derivative equations propagate the same reducible spectrum of  massless particles as the symmetric sector of the free tensionless string, as first conjectured in \cite{R2}. (See also \cite{LSM}.) 

Our work can be viewed as a continuation and an extension of the Bargmann-Wigner program, originally aimed at providing suitable wave equations for all irreducible representations of the Poincar\'e group in $D = 4$ \cite{BW}. (For a review see e.g. \cite{Buchbinder:qv}, subsection 1.8.3.) This program was later pursued for arbitrary $D$ and for the corresponding mixed-symmetry representations in a number of works, e.g. \cite{labastida,Siegel:1986zi,bb2,dmhull2,Bekaert:2003zq,Bekaert:2006ix,Skvortsov:2008vs,Alkalaev:2008gi}, while also being extended to the local Lagrangian level under various approaches --- see for instance \cite{curt, Aulakh:1986cb, Labastida:1987kw, bpt, skvort, CFMS1, clw, br, M} and references therein for massless bosonic particles with mixed-symmetry on Minkowski background. 
 
It might be worth stressing that there are two facets to this program: 
\begin{itemize}
\item[(i)] On the one hand, a goal is to provide proper covariant wave equations capable of describing  a given massless representation. (Irreducible or reducible, from our general perspective.) One of the features of the original Bargmann-Wigner equations, that we shall also stress in our approach, is that to this  specific end there is no need to invoke gauge invariance as a guiding principle in the derivation, or even to display it at all in the resulting equations. Indeed, according to the analysis that we present in section \ref{sec:2}, one can completely identify the representation of interest just assigning appropriate conditions on given $GL(D)-$tensors $\cR$, with no need for these tensors to be identified as curvatures for corresponding potentials\footnote{Strictly speaking, however, Bargmann and Wigner actually provided wave equations directly for  $SO(3,1)$-irreducible representations and never introduced any $GL(4)$ tensor. Consequently, their equations were \emph{not}  zero-trace conditions as in \eqref{traceR}.}. In this first sense, the equivalence with \eqref{laba} and \eqref{maxwell} holds at the level of the spectrum.

\item[(ii)] On the other hand, one would also like to make more direct contact with the (constrained) gauge-invariant equations \eqref{laba} and \eqref{maxwell} exploiting the gauge potential $\vf$ as the fundamental variable. As a matter of fact, the closure (``Bianchi'') conditions $d^{\, i} \cR = 0$, to be satisfied by the tensors $\cR$ as part of the system of generalised Bargmann-Wigner equations, do imply the possibility to  solve for $\cR$ in terms of $s$ exterior derivatives of a potential $\vf$ \cite{henneauxdb,bb1}. In this view, the issue becomes how to connect the resulting higher-derivative unconstrained equations to the second-order equations \eqref{laba} and \eqref{maxwell} with constrained gauge symmetry. We discuss these aspects in sections  \ref{sec:3} and \ref{sec:4}.
\end{itemize}

In relation to the second facet of our program, for the reducible Maxwell-like case of primary interest for us here it is indeed possible to exploit the Generalised Poincar\'e Lemmas \cite{Olver:1987,henneauxdb,bb1,Bekaert:2006ix} (reviewed in \cite{Bekaert:2002jn}) to first connect \eqref{divR} to an unconstrained extension of  \eqref{maxwell}, 
\be
\begin{split} \label{RMU}
d_{\, i} \, \cR \, (\vf)\,  = \, 0 \, \hskip 0.7cm  \Longleftrightarrow \hskip 0.5cm M \, = \, d^{\, i}\, d^{\, j}  \, D_{\, i j}  \, (\, \vf) \,  ,
\end{split} 
\ee
where the terms involving $D_{\, i j}  \, (\, \vf)$ emerge from the application of the corresponding cohomological analysis to the higher-derivative equation involving $\cR \, (\vf)$\ft{\label{4}As we shall recall, for the $GL(D)$-irreducible case this approach was pursued in \cite{bb2} and led to establish the equivalence
$T_{\, i j } \, \cR \, (\vf_{_Y})  \, = \, 0 \,   \Longleftrightarrow \, \cF_{_{Y}} \, = \, \tfrac{1}{2} \, {\bf Y_{\vf}}\,  d^{\, i}\, d^{\, j} \, d^{\, k} \, \cH_{\, i j k} \, (\, \vf_{_{Y}})$.
For symmetric tensors the same unconstrained ``inhomogeneous'' equations were first derived from curvatures in \cite{fs1}.}. However, in order to completely prove equivalence with \eqref{maxwell}, one still ought to discuss explicitly the issue of gauge fixing the tensors $D_{\, i j}  \, (\, \vf)$ to zero. This task, actually both for the reducible and for the irreducible cases, involves some subtleties and it is not technically straightforward to establish in general for tensors of arbitrary symmetry.

For this reason, in section \ref{sec:3} we follow a different path and propose an original argument allowing to directly establish  the equivalences:
\be
\begin{split} \label{LM}
&d_{\, i} \, \cR \, = \, 0 \, \hskip 0.65cm  \Longleftrightarrow \hskip 0.5cm M \, = \, 0 \,  , \\
&T_{\, i j } \, \cR \, = \, 0 \,  \hskip 0.5cm  \Longleftrightarrow \hskip 0.5cm \cF \, = \, 0 \, ,  \\
\end{split} 
\ee
and thus, implicitly, to also prove that the gauge fixing of the corresponding ``inhomogeneous'' unconstrained equations is indeed possible.  We collect our observations on the explicit gauge fixing in section \ref{sec:4}. 

Let us mention that, as anticipated, our results extend the scope of \eqref{traceR}, previously studied in \cite{bb2,Bekaert:2006ix} for irreps $\varphi_{_{Y}}$ of $GL(D)$, to cover the case of $GL(D)$-reducible tensors $\varphi\,$, while also yielding, as a byproduct, a relatively simple route to the degrees of freedom count of the reducible Labastida-like and of the Maxwell-like equations \eqref{laba} and \eqref{maxwell}.


\section{Reducible multiplets \`a la Bargmann-Wigner} \label{sec:2}


In this section we shall discuss Bargmann-Wigner-like equations both for irreducible and for reducible systems. By this we mean that, for given  tensors in $GL(D)$, we shall identify the conditions to be met in order for them to describe the propagation of a single-particle or of a multi-particle massless spectrum. In group theoretical terms, we shall show how to select  given representations of $O (D-2)$ or of $GL (D-2)$.

As anticipated, {\it no notion of gauge equivalence emerges at this stage}. While it is true that the $GL(D)-$tensors we shall start with may admit an interpretation as field-strengths for generalised gauge potentials, as we shall elaborate upon in the next sections, this interpretation is not needed for the purposes of this section. In this sense, the generalised Bargmann-Wigner equations provide a fully gauge-independent description of massless particles of any spin. 

Let us first discuss eq. \eqref{divR}, representing the main object of the present work. In the spirit of \cite{divR}, we would like to prove that \eqref{divR} accounts for  the degrees of freedom of the spectrum of massless particles contained in a given representation of $GL(D-2)$. To this end, we will exploit the technique used in \cite{Bekaert:2006ix,Bekaert:2003zq}. For definiteness, we shall assume this representation to be irreducible and thus to correspond to a specific tableau $Y_{\, GL(D-2)}$. The generalisation to multi-forms is straightforward. The general idea is to consider the corresponding tableau in $GL\, (D)$, but with one additional row on top, as in the following example:
\be \label{tab}
Y_{\, GL(D-2)} \, = \, \young(\hfil\hfil\hfil\hfil,\hfil\hfil\hfil,\hfil) \hskip 1cm \longrightarrow \hskip 1cm \cR_{\, GL(D)} \, = \, \young(\hfil\hfil\hfil\hfil,\hfil\hfil\hfil\hfil,\hfil\hfil\hfil,\hfil)\, 
\ee
and then to require $\cR$ to satisfy the closure and co-closure conditions
\bea 
&& d^{\, i} \, \cR \, =  0 \, , \label{B} \\
&& d_{\, i} \, \cR \, =  0 \, ,      \label{div}
\eea
where $i = 1 ,  \ldots, s$. Computing divergences of \eqref{B} leads to 
\be \label{box}
\Box \, \cR \, = \, 0 \, ,
\ee
thus implying that the representation is massless. Going to momentum space and choosing a frame where $p_{\, \mu} = (p_+, \, 0, \, \ldots, 0)$ it is then possible to observe that  \eqref{div} effectively sets to zero all components of $\cR$ with at least one ``$-$'' index in force of the equation\ft{Let us recall that in the light-cone coordinates the metric is off-diagonal along the longitudinal directions. It follows that, in the mostly-plus signature, the divergence of a vector is $p \cdot A = - p_{\, +} A_{\, -} - p_{\, -} A_{\, +} + p_{\, j} A_{\, j}$, with $j = 1, \, \ldots, \, D-2$.}
\be
p_{\, + } \, \cR_{\m^1_1 \ldots \m^1_{\ell_1 + 1}, \, \ldots, \, \m^i_1 \ldots - \ldots  \m^i_{\ell_i + 1}, 
\, \ldots, \, \m^s_1 \ldots \m^s_{\ell_s + 1}}\, = \, 0 \, ,
\ee
with indices here explicitly displayed for additional clarity.  On the other hand, whenever a given  family of indices only involves purely transverse components, then $\cR$ itself vanishes {\it tout-court}. Indeed, the ``Bianchi conditions'' \eqref{B} for the $i$th family, whenever all indices involved in the antisymmetrization take values along the $D-2$ transverse directions except for one single index along the ``$+$'' direction, reduce to the single equation
\be
p_{\, + } \, \cR_{\m^1_1 \ldots \m^1_{\ell_1 + 1}, \, \ldots, \, j^i_1 \ldots j^i_{\ell_i + 1}, \, \ldots, \, 
\m^s_1 \ldots \m^s_{\ell_s + 1}}\, = \, 0 \, .
\ee

As a result, it turns out that the only surviving components of the tensor $\cR$ for $p^{\, 2} = 0$ are those possessing only one ``$+$'' index in each family (having two or more ``$+$'' indices in the same family being forbidden by antisymmetry), while all other indices in the same family taking values along the $D-2$ transverse directions,
\be \label{irrep}
\cR_{+\, j^1_1 \ldots j^1_{\ell_1}, \, \ldots, \, +\, j^i_1 \ldots j^i_{\ell_i}, \, \ldots, \, + \,j^s_1 \ldots j^s_{\ell_s}}\, ,
\ee
thus proving that the equations \eqref{B} and \eqref{div} select from $\cR$ the components of the irrep of $GL (D-2)$ corresponding to the tableau obtained by $\cR$ itself upon removing the upper row. In these terms \eqref{B} and \eqref{div} describe the propagation of a multiplet of massless particles, identified through the branching of the $GL(D-2)-$irrep selected above in terms of its $O(D-2)$ components.

The restriction to a single irrep of $O(D-2)$, i.e. to a single massless particle, is obtained by imposing the equations \cite{bb1,bb2,Bekaert:2003zq}
\bea 
&& d^{\, i} \, \cR \, =  0 \, ,  
\label{B2} \\
&& T_{\, i j} \, \cR \, =  0 \,  , \label{T}
\eea
for $i, j = 1 ,  \ldots, s$, where in particular  the ``Bianchi conditions'', here reproduced in \eqref{B2}, are still part of the system, while \eqref{div} has been substituted by the Ricci-like equation \eqref{T} enforcing the vanishing of all the traces of $\cR\,$. Let us remark that the transversality conditions \eqref{div} are also contained in the above system, as they emerge computing traces of \eqref{B2} and using \eqref{T}. One can then go through the same steps as above and conclude that the only non-vanishing components of $\cR$ correspond to $p^{\, 2} = 0$ and are given by the traceless part of \eqref{irrep}. This selects the irrep of $O(D-2)$ formally described by the same tableau as the one identifying the $GL(D-2)-$representation previously recovered exploiting the transversality condition only.
 
Although we did not make use of it, one has to recall that due to the Generalised Poincar\'e Lemmas, it is always possible to solve the ``Bianchi conditions'' \eqref{B} and express the tensors $\cR$ as generalised curvatures for given $GL(D)-$gauge potentials $\vf$. (Or $\vf_{_Y}$, in the irreducible case.) In the next section we shall exploit this option, aiming at establishing the links between the higher-derivative equations for $\vf$ resulting from \eqref{div} and \eqref{T} and the second-order equations for the corresponding particle content \eqref{maxwell} and \eqref{laba}, respectively.



\section{Maxwell-like equations from curvatures} \label{sec:3}


In the present section we shall consider the tensors $\cR$ as generalised curvatures for corresponding potentials: $\cR = \cR\, (\vf)$. Their construction can be illustrated in several ways; one possibility is to motivate their explicit form as providing the solution to the ``Bianchi conditions'' 
\be \label{B3}
d^{\, i} \, \cR \, = \, 0 \, ,
\ee
obtained through the Generalised Poincar\'e Lemmas, with the following outcome: if $\vf_{_{Y}}$ is an irreducible tensor of $GL (D)$ described by a tableau with $s$ columns, then the general solution to  \eqref{B3} determines its corresponding curvature as  the $s$th derivative combination
\be \label{R}
\cR \, (\vf_{_{Y}}) \, = \, d^{\, 1} \, d^{\, 2} \, \cdots \, d^{\, s} \, \vf_{_{Y}} \, . 
\ee
One can then check that the tensor \eqref{R} is identically gauge-invariant under the $Y_{GL(D)}$-projected version of \eqref{gauge}, satisfies the Bianchi identities \eqref{B} and corresponds to the irrep of $GL(D)$ obtained from $Y$ by adding an extra row on top of it, as pictorially suggested by the following example:
\be \label{tab}
\vf_{_{Y}} \, = \, \young(\hfil\hfil\hfil\hfil,\hfil\hfil\hfil,\hfil) \hskip 1cm \longrightarrow \hskip 1cm \cR \, (\vf_{_{Y}}) \, = \, \young(\pr\pr\pr\pr,\hfil\hfil\hfil\hfil,\hfil\hfil\hfil,\hfil)\, 
\ee
Similarly, starting from a multi-form potential $\vf$ one can obtain the corresponding curvature by computing the tensor-product of the various forms entering its definition, each being differentiated once,
\be \label{multifR}
\vf \, = \, \mbox{\small$\young(1,\tvdots,\1)\, \otimes \, \young(1,\tvdots,\2) \, \otimes \, \dots \, \otimes \, \young(1,\tvdots,\ss)$}\, 
\hskip 1cm \longrightarrow \hskip 1cm \cR \, (\vf) \, = \, \mbox{\small$\young(\pr,1,\tvdots,\1)\, \otimes \, \young(\pr,1,\tvdots,\2) \, \otimes \, \dots  \, \otimes \, \young(\pr,1,\tvdots,\ss)$}\,
\ee
thus obtaining an expression for $\cR \, (\vf)$ equivalent to \eqref{R}. 

The tensors \eqref{R} appear as very natural objects in the gauge theory involving the field $\vf$, and indeed we already observed that the Ricci-like equations $T_{\, i j } \, \cR = 0$ do play a central role both for low-spin and for high-spin theories.  Our main objective here is to connect curvature tensors with the second-order equations for reducible multiplets \eqref{maxwell}. However, since our analysis directly applies to the Labastida case as well,  we shall effectively establish the following two equivalences:
\be
\begin{split} \label{LM1}
&d_{\, i} \, \cR  \, (\vf)\, = \, 0 \, \hskip 0.65cm  \Longleftrightarrow \hskip 0.5cm M \, = \, 0 \,  , \\
&T_{\, i j } \, \cR  \, (\vf)\, = \, 0 \,  \hskip 0.5cm  \Longleftrightarrow \hskip 0.5cm \cF \, = \, 0 \, ,
\end{split} 
\ee
thereby also extending the results of \cite{bb2,Bekaert:2006ix} to the case of {\it reducible} $GL(D)-$potentials. In the remainder of the section we shall consider multi-form fields $\vf$.  The restriction to tableaux in $GL(D)$  is essentially immediate. 

The general idea of our argument is to exploit the results of the previous section and solve for $\vf$ in \eqref{R}, so as to extract information on its non-vanishing components. To this end, let us first consider the transversality conditions
\be \label{D}
d_{\, i} \, \cR \, (\vf)\, = \, 0 \, ,
\ee
and let us recall that the surviving components in $\cR (\vf)$ are those of the form \eqref{irrep}, corresponding to $p^{\, 2} = 0$. Thus, in the frame where only $p_{\, +}$ is different from zero, we can immediately deduce consequences for two classes of components of $\vf$:
\begin{description}
 \item{-} purely transverse components:
\be \label{trans}
\vf _{j^1_1 \ldots j^1_{\ell_1}, \, \ldots, \, j^i_1 \ldots j^i_{\ell_i}, \, \ldots, \, j^s_1 \ldots j^s_{\ell_s}} 
\, = \, 
\fr{1}{(p_{\, +})^{\, s}}\,  \cR_{+ \, j^1_1 \ldots j^1_{\ell_1}, \, \ldots, \, +\, j^i_1 \ldots j^i_{\ell_i}, \, 
\ldots, \, + \, j^s_1 \ldots j^s_{\ell_s}}\, .     \\
\ee
They are gauge invariant in the chosen frame and are not constrained by closure and co-closure conditions on $\cR \, (\vf)$. Thus, they represent propagating components of $\vf$  in \eqref{D} satisfying the D'Alembert equation;
\item{-} components with mixed transverse and  ``$-$'' indices:
\be
\begin{split}
\vf _{j^1_1 \ldots j^1_{\ell_1}, \, \ldots, \, j^i_1 \ldots \, - \, \ldots j^i_{\ell_i}, \, \ldots, \, j^s_1 \ldots j^s_{\ell_s}} \, = \,  
\fr{1}{(p_{\, +})^{\, s}}\,  \cR_{+ \, j^1_1 \ldots j^1_{\ell_1}, \, \ldots, \, +\, j^i_1 \ldots\, - \, \ldots j^i_{\ell_i}, \, \ldots, \, + \, j^s_1 \ldots j^s_{\ell_s}}\, = \, 0 \, .
\end{split}
\ee
(``$-$'' indices can be present in more than one family, of course.) They are also gauge invariant, but vanish due to the co-closure conditions \eqref{D}.
\end{description}
Clearly, from the equations \eqref{D} we cannot obtain information on the components of $\vf$ that are still gauge-dependent even in the chosen frame; these are all the components possessing at least one index along the ``$+$'' direction, all other indices belonging to the same 
family being either ``$-$'' or transverse (we shall denote them collectively by capital $\j$'s), while staying arbitrary for the other families:
\be \label{plus}
\vf _{\m^1_1 \ldots \m^1_{\ell_1}, \, \ldots, \, \j^i_1 \ldots \, + \, \ldots \j^i_{\ell_i}, \, \ldots, \, 
\m^s_1 \ldots \m^s_{\ell_s}} \, .
\ee
This is the point where gauge invariance of the curvatures $\cR \, (\vf)$ under \eqref{gauge} plays its role, since in force of the unconstrained nature of the gauge symmetry of \eqref{R} we are in the position to eliminate all components of the form \eqref{plus} {\it performing a complete gauge fixing}. As a result, only the components identified in \eqref{trans} represent propagating degrees of freedom, all the others being either vanishing or pure gauge.

Let us be more precise. The gauge transformations \eqref{gauge} read, in components:
\be
\begin{split}
\delta_{\, \Lambda}\, \varphi_{\m^1_1 \ldots \m^1_{\ell_1}, \, \m^2_1 \ldots \m^2_{\ell_2}, \, \m^3_1 \ldots \m^3_{\ell_3}\, \ldots} \, 
= & \, \partial_{\m^1_1}\, \Lambda_{(1)}{}_{\m^1_2 \ldots \m^1_{\ell_1}, \, \m^2_1 \ldots \m^2_{\ell_2}, \, \m^3_1 \ldots \m^3_{\ell_3}\, \ldots} \\
+ & \, \partial_{\m^2_1}\, \Lambda_{(2)}{}_{\m^1_1 \ldots \m^1_{\ell_1}, \, \m^2_2 \ldots \m^2_{\ell_2}, \, \m^3_1 \ldots \m^3_{\ell_3}\, \ldots} \\
+ &  \, \partial_{\m^3_1}\, \Lambda_{(3)}{}_{\m^1_1 \ldots \m^1_{\ell_1}, \, \m^2_1 \ldots \m^2_{\ell_2}, \, \m^3_2 \ldots \m^3_{\ell_3}\, \ldots}\, + \ldots
\end{split}
\ee
where we treat the reducible case. We proceed with the gauge fixing as follows:
\begin{itemize}
\item[(1)] One fixes $\varphi_{+ \j^1_2 \ldots \j^1_{\ell_1}, \, \m^2_1 \ldots \m^2_{\ell_2}, \, \m^3_1 \ldots \m^3_{\ell_3}\, \ldots}$ to zero by using 
$\Lambda_{(1)}{}_{\j^1_2 \ldots \j^1_{\ell_1}, \, \m^2_1 \ldots \m^2_{\ell_2}, \, \m^3_1 \ldots \m^3_{\ell_3}\, \ldots}$. We note that the indices 
$ \m^2_1 \ldots \m^2_{\ell_2}, \, \m^3_1 \ldots \m^3_{\ell_3}\, \ldots$ are left totally unspecified, so that some of them can be ``$+$";
\item[(2)] One sets $\varphi_{\j^1_1 \ldots \j^1_{\ell_1}, \, + \j^2_2 \ldots \j^2_{\ell_2}, \, \m^3_1 \ldots \m^3_{\ell_3}\, \ldots}$ to zero by using
$\Lambda_{(2)}{}_{\j^1_1 \ldots \j^1_{\ell_1}, \, \j^2_2 \ldots \j^2_{\ell_2}, \, \m^3_1 \ldots \m^3_{\ell_3}\, \ldots}$. Note that this gauge fixing does not affect  the previous one;
\item[(3)] One sets $\varphi_{\j^1_1 \ldots \j^1_{\ell_1}, \, \j^2_1 \ldots \j^2_{\ell_2}, \, + \j^3_2 \ldots \j^3_{\ell_3}\, \ldots}$ to zero by using
$\Lambda_{(3)}{}_{\j^1_1 \ldots \j^1_{\ell_1}, \, \j^2_1 \ldots \j^2_{\ell_2}, \, \j^3_2 \ldots \j^3_{\ell_3}\, \ldots}$, {\textit{etc}}.
\end{itemize}
Now, all the components of the gauge field are either zero by virtue of the field equations and gauge fixings, or are expressed in terms of the non-vanishing components of the curvature tensor, so we have achieved a complete gauge fixing.

In the case of the Ricci-like equation $T_{\, ij} \cR$ = 0 two differences are to be taken into account: first, as already mentioned, by virtue of the Bianchi identity \eqref{B} one can derive the vanishing of the divergences as a consequence. Second,  the surviving components of $\vf$  will define a {\it traceless tensor} in the transverse indices. 

We can summarise our findings in the following scheme, where one should keep in mind that our conclusions hold in the gauge where the components \eqref{plus} are set to zero:
\be \label{reduction}
\begin{split} 
& \cR \, (\vf) \, = \, d^{\, 1} \cdots \, d^{\, s} \, \vf \, \longrightarrow \hskip 0.1cm d^{\, i} \, \cR \, (\vf) \, \equiv \, 0 \, , \\
&d_{\, i} \, \cR \, (\vf)\, = \, 0 \, \hskip 1,25cm  \longrightarrow \hskip 0.1cm 
\begin{cases}
\vf_{\ldots, \, - \, \ldots} \, = \, 0\, , \\
\vf_{\ldots, \, + \, \ldots} \, = \, 0\, , \\
p^{\, 2} \, \vf_{\ldots, \, j \, \ldots} \, = \, 0 \, , \\
\end{cases}
\longrightarrow \hskip 0.5cm
\vf \hskip 0.5cm  \mbox{purely transverse}
, \\
&T_{\, i j } \, \cR \, (\vf) \, = \, 0 \,  \hskip 1.1cm  \longrightarrow \hskip 0.1cm
\begin{cases}
d_{\, i} \, \cR \, = \, 0 \\
T_{\, i j} \vf \, = \, 0  \, 
\end{cases}
\hskip 0,45cm \longrightarrow \hskip 0.5cm
\vf \hskip 0.5cm  \mbox{transverse and traceless.}     \\
\end{split} 
\ee
Similarly, if $\vf_{_{Y}}$ takes value in a given irrep $Y$ of $GL (D)$, the equations obtained setting to zero the divergences or the traces of the curvatures \eqref{R} propagate only {\it transverse} components of $\vf_{_{Y}}$, corresponding to irreps of $GL (D-2)$ in the former case or of $O (D-2)$ in the latter.

In order to make contact with the Maxwell-like equation \eqref{maxwell} and with the Labastida equation \eqref{laba}, and thus to complete our proof of the equivalences \eqref{LM1}, one further step is needed consisting in explicitly computing divergences and traces of $\cR \, (\vf)$,  so as to show that the following relations hold:
\be
\begin{split} \label{O}
&d_{\, i} \, \cR \, (\vf) \, = \, \cO^M{}_i \, (d) \, M , \\
&T_{\, i j } \, \cR \, (\vf) \, = \, \cO^F{}_{ij}\, (d) \, \cF \, ,  
\end{split} 
\ee
where $\cO^M{}_i \, (d)$ and $\cO^F{}_{ij}\, (d)$ are linear and homogeneous differential operators built out of  the exterior derivatives. For instance, computing the divergence of \eqref{R} in the first family we obtain
\be \label{divRM}
\begin{split}
d_{\, 1} \, \cR \, (\vf)\, & = \,  d_{\, 1}  d^{\, 1} \, d^{\, 2} \, \cdots \, d^{\, s} \, \vf, \\
& = (\Box \, - \, d^{\, 1}  d_{\, 1}) \, d^{\, 2} \, \cdots \, d^{\, s} \, \vf \\
& = d^{\, 2} \, \cdots \, d^{\, s} \, (\Box \, - \, d^{\, 1}  d_{\, 1}) \, \vf \\
& = d^{\, 2} \, \cdots \, d^{\, s} \, (\Box \, - \, d^{\, i}  d_{\, i}) \, \vf \, ,
\end{split}
\ee
where in particular the last step holds due the vanishing of each product $d^{\, k} \, d^{\, k}$ (no summation implicit), allowing to reconstruct the full Maxwell operator acting on $\vf$. Similarly for the second equation in \eqref{O}, selecting for instance the trace in first and second family,  
one gets \cite{bb2,Bekaert:2006ix}
\be
T_{\, 12 } \, \cR  \, (\vf) \, = \,  d^{\, 3} \, \cdots \, d^{\, s} \, \cF \, .
\ee

Equations \eqref{O} make clear that $M = 0$ and $\cF = 0$ do provide solutions to  $d_{\, i} \, \cR \, = \, 0$ and $T_{\, i j } \, \cR \, = \, 0$. The nontrivial part of the argument, in general, is to show that there are no other solutions, in spite of the different derivative order of the two sets of equations.

However, after our discussion also this further step is by now almost transparent. Indeed, it is sufficient to explicitly observe that field configurations satisfying the second set of conditions in \eqref{reduction} {\it automatically satisfy $M = 0$},  due to the separate cancellation of the two terms in \eqref{maxwell}. Similarly, if to the above conditions we add the tracelessness of $\vf$, by the same reasoning we can conclude that those configurations {\it automatically satisfy $\cF = 0$}, thus establishing \eqref{LM1}. The fact that the equations in the left column of \eqref{reduction} display a bigger gauge invariance than \eqref{laba} and \eqref{maxwell}, is a manifestation of the existence of a set of  equivalences slightly different from \eqref{LM1}, anticipated in the introduction, involving gauge invariant completions of the tensors $M$ and $\cF$ and from which \eqref{LM1} actually arise as the result of a partial gauge fixing. We discuss this option in the next section.

We could even draw our conclusions about \eqref{LM1} by an alternative line of reasoning. Indeed, for $GL ( D)-$irreducible potentials $\vf_{_{Y}}$, the linear space of solutions to the Labastida equation \eqref{laba} is obviously a representation of $ISO(D-1,1)$ since the equation is covariant. We already observed that, in force of \eqref{O}, the same space of solutions has to be contained in that of the equation $T_{\, ij} \,{\cal R} \, (\vf_{_Y})=0\,$. The latter, on the other hand, according to our analysis of section \ref{sec:2}, actually is $ISO(D-1,1)-$\emph{irreducible}, thus implying that the two spaces have to coincide. The argument applies equally well to multi-forms, upon projecting all the equations onto the corresponding irreducible spaces.

For the Maxwell-like equations \eqref{maxwell}, on the other hand, a complete counting of degrees of freedom was performed in \cite{M}. Comparing with the results of section \ref{sec:2} and taking into account \eqref{O} would already suffice to establish the first equivalence in \eqref{LM1}.  
 
Finally, let us also observe that the argument presented in this section, in conjunction with our results of section \ref{sec:2}, provides a relatively simple route to the explicit degrees of freedom count of both eqs. \eqref{laba} and \eqref{maxwell}. 


\section{On the gauge fixing of curvature equations} \label{sec:4}


Our approach in the previous sections largely exploited the possibility to choose a preferred Lorentz frame where the only non-vanishing component of the momentum $p_{\, \m}$ was $p_{\, +}$, leading to relevant technical simplifications. In this section we would like to comment on the covariant meaning of our procedure, also in order to establish contact with the papers \cite{bb2, Bekaert:2006ix} and \cite{R2, divR, divRbis}, of which our present work represents a natural continuation and completion.

An alternative path to discussing the equivalences \eqref{LM1} would proceed  in two steps. In the first step, starting with \eqref{O} one would first solve covariantly the equations 
\be
\begin{split} \label{OMF}
&\cO^M{}_i \, (d)\, M \, = \, 0 \, , \\
&\cO^F{}_{ij}\, (d) \, \cF \,  = \, 0 \, ,  \\
\end{split} 
\ee
essentially looking for the general form of the kernels of the operators $\cO^M{}_i \, (d)$ and $\cO^F{}_{ij}\, (d) $. Exploiting to this end the general results of \cite{henneauxdb} and \cite{bb1}, one can then show that for multi-forms $\vf\,$, a weak form of \eqref{LM1} actually holds, where on the right-hand side gauge-invariant completions of the Maxwell-like and of the Labastida equations would appear\ft{See also \cite{BBI} for a related discussion in the context of quantum forms on K¨ahler spaces.}
\be
\begin{split} \label{LMweak}
&d_{\, i} \, \cR \, (\vf) \, = \, 0 \, \hskip 0.6cm  \Longleftrightarrow \hskip 0.5cm M \, =  \,d^{\,  i} \, d^{\, j}  \, D_{\, i j} \, (\vf) \, ,\\
&T_{\, i j } \, \cR \, (\vf) = \, 0 \,  \hskip 0.5cm  \Longleftrightarrow \hskip 0.5cm \cF \, = \, \tfrac{1}{2} \, d^{\,  i} \, d^{\, j} \, d^{\, k} \, \cH_{\, i j k} \, (\vf)   \, ,
\end{split} 
\ee
while the corresponding relations for $GL (D)-$irreps would involve suitable projectors ${\bf Y_{\vf}}$ acting on the compensator terms\ft{It might be worth recalling  that the operators entering the definitions of $M$ and of $\cF$ automatically enforce the $GL(D)-$projection, in the sense that when computed on irreducible potentials $\vf_{_{Y}}\,$, both $M$ and $\cF$ define $GL(D)-$irreducible kinetic tensors \cite{Bekaert:2006ix, CFMS1, M}.} 
$d^{\,  i} \, d^{\, j}  \, D_{\, i j} \, (\vf) $ and $d^{\,  i} \, d^{\, j} \, d^{\, k} \, \cH_{\, i j k} \, (\vf)$.  

Let us stress that, in force of the general discussion we put forward in section \ref{sec:2}, the existence of the equivalences \eqref{LMweak} allows to directly conclude that the wave equations to the r.h.s. of \eqref{LMweak} do describe the correct massless representations of interest, regardless the possibility to consistently truncate them to  \eqref{maxwell} and \eqref{laba}. Indeed, if we did not have alternative arguments to confirm the particle content of the ``ordinary'' wave equations \eqref{laba} and \eqref{maxwell}, we should rather conclude that the equations in \eqref{LMweak} are the correct ones, to the purpose of describing given massless representations, and leave it open the issue about the status of their constrained counterparts. Starting from \eqref{LMweak}, in order to fully establish the equivalence \eqref{LM} at the covariant level, one should discuss the possibility of eliminating the tensors  $D_{\, i j} \, (\vf)$ and $\cH_{\, i j k} \, (\vf)$ by means of a suitable partial gauge fixing. However, this very step involves some subtleties that are not easily sorted out in the general case.

On the one hand, even without computing the explicit form of $\cH_{\, i j k} \, (\vf)$ and $D_{\, i j} \, (\vf)$ one might infer by consistency the form of their gauge transformations 
\be
\begin{split} \label{HDgauge}
&\d \,  \, \cH_{\, i j k} \, (\vf)   \, = \, \tfrac{1}{3}\,  T_{\, (ij} \, \L_{\, k)}\, ,     \\
&\d  \, D_{\, i j} \, (\vf) \, = \, \tfrac{1}{2} \, d_{\, (i} \L_{\, j)}\, .
\end{split} 
\ee
However, the transformations  in \eqref{HDgauge} {\it are not all independent} and thus would not allow a complete elimination of the corresponding tensors $\cH_{\, i j k}$ and $D_{\, i j}$ in general, as if they were independent Stueckelberg fields\ft{See Appendix \ref{A2} for an explicit example.}. (One notable exception being the case of fully symmetric tensors \cite{fs3, M}.) As originally observed in \cite{CFMS1} for the first of \eqref{HDgauge}, the compensators could indeed be gauged away if one could prove (or assume, if they were to be regarded as additional independent fields) that their structure is of the form
\be \label{comp}
\cH_{\, ijk} \, \sim \, T_{\, (i j } \Th_{\, k)}\, ,
\ee
with $\Th_{\, k}$ s.t. $\d \, \Th_{\, k} \, = \, \L_{\, k}$. On the other hand, to compute in general the form of $\Th_{\, k} \, (\vf)$ stemming from the application of the  Generalised Poincar\'e Lemmas to the equation $T_{\, i j }  \cR (\vf) = 0$ would be technically rather involved. For the Maxwell-like case, moreover, in addition to similar considerations one would face further subtleties related to the existence of gauge transformations that are only effective for $p^{\, 2} = 0$ and that are crucial to discuss the spectrum of the equations \eqref{maxwell}\ft{See section $3.1.2$ of \cite{M}.}.

While it would still be possible in principle to proceed along this path, and so to provide a complete covariant demonstration of the equivalences \eqref{LM}, we would like to stress that the proof that we proposed in the previous section allows not only to completely bypass the issue of gauge fixing of the tensors $\cH_{\, i j k} \, (\vf)$ and $D_{\, i j} \, (\vf)$ in \eqref {LMweak}, but also to actually show, indirectly, that such a gauge fixing is indeed possible, thus dispensing us with the need to compute their effective form as functions of $\vf$.

Let us also comment on the issue of double-trace constraints present in the Fronsdal-Labastida theory \cite{fronsdal, labastida}.  The trace constraints on the gauge parameters in \eqref{laba} imply that some combination of double traces of $\vf$ are gauge invariant. Still, one needs to get rid of them on-shell, so as to avoid the propagation of unwanted degrees of freedom. The solution to this issue usually considered in the literature is to assume that those gauge-invariant combinations have to vanish identically:
\be \label{dtrace}
T_{\, (ij} \, T_{\, kl)} \, \vf \, \equiv \, 0 \, .
\ee
However, as first mentioned in \cite{dariocrete} for the case of symmetric tensors, the non-Lagrangian equations $\cF = 0$ are indeed strong enough to imply by themselves the vanishing of double traces, so that there is no need to enforce \eqref{dtrace} as independent conditions whenever the equations $\cF = 0$ hold\ft{The reason is the existence of the ``contracted Bianchi identity'' $\prd \cF \, - \, \frac{1}{2} \, \pr \cF^{\, \pe} \, \equiv \, - \, \fr{3}{2} \, \pr^{\, 3} \, \vf^{\, \pe\pe}$, where $\cF^{\, \pe}$ and $\vf^{\, \pe \pe}$ denote the trace of $\cF$ and the double trace of $\vf$, respectively.}. For mixed-symmetry fields the same issue was discussed in section 2.1.2 of \cite{M}, where it was observed that the Bianchi identities satisfied by $\cF$ allow to conclude that weaker double-trace constraints would indeed suffice to keep consistency, as long as \eqref{laba} hold.\footnote{In the mixed-symmetry case, one could think of applying the Generalised Poincar\'e Lemma to the Bianchi identity 
$d_{\, i}\, \cF - \frac{1}{2}\,d^{\, j}\, T_{ij}\, \cF \equiv -\, \frac{1}{12} \, d^{\, j}d^{\, k}d^{\, l}\, T_{\, (ij} \, T_{\, kl)}\, \vf$
\cite{CFMS1}, so as to directly derive \eqref{dtrace} from \eqref{laba}. However, setting to zero the right-hand side of this identity is not a cocycle condition to which the Generalised Poincar\'e Lemma can be applied.}

Here we are in the position of completing the argument and of actually concluding that even for arbitrary mixed-symmetry fields the Labastida equations \eqref{laba} are strong enough to  imply \eqref{dtrace} among their consequences. This result stems from our analysis performed in sections \ref{sec:3} and \ref{sec:4}. Indeed, the second equivalence in \eqref{LM1} also implies that the double traces \eqref{dtrace} are to vanish on shell (otherwise the equation $\cF = 0$ would propagate more degrees of freedom than its geometric counterpart $T_{\, ij} \cR = 0$), as also manifest in the gauge where the components \eqref{plus} are set to zero.

This being said, the double-trace constraints are necessary if one wishes to derive \eqref{laba} as a consequence of a variational equation. Already for symmetric tensors indeed it is possible to show that, in the absence of auxiliary fields, the gauge-invariant equations of motion derived from the Fronsdal Lagrangian\ft{Assuming traceless gauge parameters the Fronsdal Lagrangian, written as a sum of bilinears in the potential $\vf$, is gauge invariant regardless the field being or not doubly-traceless.} cannot be strong enough to enforce \eqref{dtrace} by themselves. For this reason, in order to maintain consistency, in the sense of deriving $\cF = 0$ from the Lagrangian equations, one has to perform the variation of the action over a space of fields constrained as in \eqref{dtrace} \cite{DFun}.

For the Maxwell-like equations \eqref{maxwell} a similar issue arises, since the differential conditions on the gauge parameters imply that a number of double divergences of $\vf$ are gauge-invariant, thus posing a problem concerning their elimination. We refer the reader to \cite{M}, where the issue is discussed in detail, and only mention here that the Maxwell-like operator in \eqref{maxwell} displays enough gauge symmetry so as to remove all components of the double divergences that are not set to zero by the equations of motion themselves. With hindsight, the arguments we present in this work also provide an alternative proof to the same effect.


\section{Conclusions}


In this paper we discussed the higher-spin equations obtained setting to zero the divergences or the traces of curvature tensors of arbitrary symmetry types. 

Transversality conditions emerge as consequences of the Bianchi identities whenever curvatures are imposed to be traceless, as needed to the goal of describing the propagation of individual particles with a given spin. What we showed in this work, following the results of \cite{divR} for the case of symmetric tensors, is that even imposing transversality conditions {\it alone} leads to a consistent description of massless  particles, propagating in this case in {\it reducible multiplets}  organized as representations of $GL(D-2)$.  

A notable feature of the derivation lies in the equivalence, up to a suitable partial gauge fixing, between the higher-derivative equations obtained setting the divergences of the curvatures to zero and the second-order wave equations for the same gauge potentials given in \eqref{maxwell}. One possible key to interpret such an unexpected property lies in our analysis of section \ref{sec:2}. There we showed how to extract the particle content of closure and transversality conditions with no need to interpret the tensors under scrutiny as curvatures for gauge potentials. In this approach the resulting wave equations are standard, second-order ones (albeit for different tensors) and in this sense one could expect that their content does not change when the same tensors are solved for as higher-derivative curvatures for gauge potentials. 

One drawback of equations \eqref{traceR} and \eqref{divR} is that they are intrinsically non-Lagrangian, while standard actions for the same particle content typically do not involve curvatures and do not display full gauge invariance \cite{fronsdal, labastida, M}. Lagrangians involving curvature tensors analogous to \eqref{R}, on the other hand, typically involve non-local terms \cite{fs1, Bekaert:2006ix, FMS, R2}. At least in some cases, the latter can be interpreted as due to the integration over unphysical field components needed off-shell to enforce full gauge invariance \cite{dariocrete, R2}. In this sense, the way non-localities appear closely resembles the emergence of non-local terms in the Polyakov action after the integration over the pure-gauge (even at the quantum level, in the critical dimension) Liouville mode \cite{Polyakov:1981rd}. 

Our analysis covers all possible types of bosonic massless particles with finite spin in arbitrary dimension $D$, covariantly described in terms of  the ``standard'' representation for gauge fields first proposed by Labastida. On the other hand, we did not discuss the infinitely-many different 
$GL(D)-$irreducible field representations arising upon dualisations \cite{Boulanger:2012df,Boulanger:2012mq}, which might represent a possible direction to explore. More generally, it would be interesting to study the issue of duality from the perspective of reducible  particle models. 

Reducible systems of fermionic massless particles were considered in \cite{triplets2, fs2, st, SV, FT, R2} for symmetric spinor-tensors. Fermions with mixed symmetry were discussed in \cite{labaferm} and first given a full Lagrangian formulation in \cite{CFMS2}. See also  \cite{Skvortsov:2010nh, Reshetnyak:2012ec}. However, the generalization of \eqref{divR} to the case of fermions does not appear to be straightforward: while setting to zero a single $\g-$trace of fermionic curvatures provides a counterpart of \eqref{traceR}, and as such selects the polarisations of a single particle \cite{BBAST}, computing their divergences would more directly connect  to less natural second-order wave equations. We leave to future work a more accurate analysis of this issue, as well as of the (A)dS deformation of the results illustrated in the present work. 



\section*{Acknowledgments}

We would like to thank the Galileo Galilei Institute for Theoretical Physics for the hospitality  during the workshop ``Higher Spins, Strings and Duality'', and the INFN for partial support. The research of X.B.  was partially supported by the Russian Science Foundation grant 14-42-00047 in association  with Lebedev Physical Institute. The work of N.B. was partially supported by a contract  ``Actions de Recherche concert\'ees -- Communaut\'e fran\c{c}aise de  Belgique''  AUWB-2010-10/15-UMONS-1. N.B. wants to thank the Scuola Normale Superiore, Pisa, for inviting him for a  collaboration stay during which part of this work was done. D.F. is grateful to A. Campoleoni for useful discussions. In addition, D.F. would like to thank the Service de M\'ecanique et Gravitation at Mons University for the invitation to the workshop ``On Higher-Spin Gravities and Related Topics'', September 4-6, 2012, in the occasion of which this project was started, and  also wishes to express his gratitude to CNRS-UMR7350 LMPT Tours for partial support and for  the kind hospitality extended to him during the completion of this work.


\begin{appendix}


\section{Notation and conventions} \label{A1}


The basic objects in our work are tensors in $GL(D)$, either reducible or irreducible. As a rule, we denote the former with $\vf$ and the latter with $\vf_{_{Y}}$, hinting to the possibility of associating with a given irrep a corresponding Young tableau $Y$. Generally speaking, such tensors possess a number of ``families'' of indices with explicit symmetry or anti-symmetry properties within each family, depending on the chosen basis. Thus, one refers to reducible $GL(D)-$tensors as {\it multi-symmetric tensors} whenever the various families correspond to tensor-products of symmetric tensors
\be \label{multis}
\vf \, \equiv \, \varphi_{(\mu_1 \cdots\, \mu_{\1}),\,(\nu_1 \cdots\, \nu_{\2}),\,\cdots} \, = \, \mbox{\small$\young(1\cdots \1)\, \otimes \, \young(1\cdots \2) \, \otimes \, \dots $}\, ,
\ee
with parentheses to signify symmetrisation with no additional overall factors, or, differently, to {\it multi-forms} in case the reducible $GL(D)-$representation arises from products of forms of various degrees
\be \label{multif}
\vf \, \equiv \, \varphi_{[\mu_1 \cdots\, \mu_{\1}],\,[\nu_1 \cdots\, \nu_{\ell_{2}}],\,\cdots} \, = \, 
\mbox{\small$\young(1,\tvdots,\1)\, \otimes \, \young(1,\tvdots,\2) \, \otimes \, \dots $}\, ,
\ee
with antisymmetrization of indices (with no overall factors) denoted by square brackets.
Similarly, {\it irreducible} tensors in $GL(D)$ can be described in terms of Young 
diagrams, e.g.
\be
\vf_{_{Y}} \, = \, \young(\hfil\hfil\hfil\hfil,\hfil\hfil\hfil,\hfil) \, ,
\ee
where either manifest symmetry along rows or manifest antisymmetry along columns can be enforced. Irreps contained in the tensor product \eqref{multis} will be of the first type, while the Young tableaux providing the decomposition of the multiform \eqref{multif} will be of the second type. Both conventions are widely used in the literature and they turn out to display specific advantages or drawbacks depending on the problem under consideration. We refer the reader to the appendices of \cite{CFMS1, A} for more details on the symmetric convention and to that of \cite{Bekaert:2006ix} for the multi-forms and the use of the antisymmetric basis. In particular, in order to construct and to study the properties of generalised curvatures the antisymmetric convention turns out to be more convenient, so that in this paper our tensors are always meant to be multiforms in the reducible case, or anyway to display explicit antisymmetry along columns in the irreducible case. However, in Appendix \ref{A2} we exploit multi-symmetric tensors to provide an example of gauge-invariant combination of traces of compensators arising from the second of \eqref{HDgauge}.

In order to keep our formulas readable usually we do not display space-time indices, while we introduce family indices denoted by small-case Latin letters. We are thus able to identify tensors carrying a different number of  indices in some sets as compared to the basic field $\vf$, while also keeping track of index-reshuffling among different families, according to the position (up or down) of the family index. 

For instance, the gauge parameters are denoted by $\L_{\, i}$, with a {\it lower} index $i$ to indicate that they carry one index less than the gauge field $\vf$ in the $i$th family.  Differently, an {\it exterior derivative} carrying a space-time index to be anti-symmetrised with indices belonging to the $i$th group is denoted by the usual symbol $d$ in conjunction with an {\it upper} index $i$: 
\be \label{notation}
d^{\,i} \, \vf \, :=\,  \, \pr_{\,[\,\m^i{}_1|} \, \vf_{\,\cdots \,,\, | \, \m^i{}_2 \, \cdots \, \m^i{}_{\ell_i+1} ] \,,\, \cdots} \, ,
\ee
where the antisymmetrization involves the minimal number of terms, with no overall factors.  The Einstein convention for summing over pairs of indices is used throughout. A notable example is the gauge transformation \eqref{gauge} of $\vf$ ,
\be
\d \, \vf \, = \, d^{\, i} \, \L_{\, i}\, ,
\ee
given by a sum of exterior derivatives,  each for any of the families of $\vf$. 

In a similar spirit, for a {\it divergence} contracting an index in the $i$th group we use the notation
\be
d_{\,i} \, \vf \, := \; \; \pr^{\,\l} \, \vf_{\,\cdots \,,\, \l \, \m^i{}_1 \, \cdots \, \m^i{}_{\ell_i-1} \,,\, \cdots} \, ,
\ee
while {traces}, contracting one index in family $i$ with one index in family $j$ (with $i \neq j$ in the antisymmetric convention), are denoted by
\be
T_{ij} \, \vf \, := \;\;\vf_{\ldots, \, \phantom{\m^i_1}}{}^{\!\!\!\l}{}_{\m^i_2 \, \ldots \, \m^i_{\ell_i},\, \ldots, \, \l \, \m^j_2 \ldots \, \m^j_{\ell_j}, \, \ldots}\, .
\ee
In order to manipulate the various formulae efficiently one has to keep in mind a few relations (for more details, see \cite{Bekaert:2006ix}) relevant for instance in computing \eqref{divRM}: 
\begin{align}
& d_{\, i}\, d^{\, j}\, = \,(-1)^{\delta_{ij}}\, d^{\, j} d_{\, i}\,+\, \Box\,\delta_i^j \, , \\
& d^{\, i}\, d^{\, j}\, = \,(-1)^{\delta_{ij}}\, d^{\, j} d^{\, i} \, , \\
\,& [\, T_{\, ij } \,,\, d^{\, k}\,] \,=\, 2 \, \d^{\, k}{}_{(i} \, d_{\, j)} \, .
\end{align}
%


\section{A gauge-invariant combination of compensators} \label{A2}


Consider a $\{4, 3\}$ multi-symmetric tensor field 
$\vf = {\tiny \young(\hfil\hfil\hfil\hfil) \otimes \young(\hfil\hfil\hfil)}$\, . Its gauge transformation involves two reducible parameters 
$\L_{\, 1} = {\tiny \young(\hfil\hfil\hfil) \otimes \young(\hfil\hfil\hfil)}$ and 
$\L_{\,2} = {\tiny \young(\hfil\hfil\hfil\hfil) \otimes \young(\hfil\hfil)}$\, , and in explicit notation would look
\be
\d \, \vf_{\, \m_1 \m_2 \m_3 \m_4, \, \n_1 \n_2 \n_3} \, = \, \pr_{\, (\m_1} \, \L^{(1)}{}_{\, \m_2 \m_3 \m_4), \, \n_1 \n_2 \n_3}\, + \, 
\pr_{\, (\n_1|}\, \L^{(2)}{}_{\, \m_1 \m_2 \m_3 \m_4,| \, \n_2 \n_3)}\, .
\ee
Correspondingly, in the unconstrained extension of the Labastida equation for $\vf$ there would appear four possible compensator structures $\cH_{\, 111}, \, \cH_{\, 112},\, \cH_{\, 122}, \, \cH_{\, 222}$, s.t.
\be
\begin{split}
&\d \, \cH_{\, 111} \, = \, T_{\, 11} \, \L_{\, 1} \, ,\\
&\d \, \cH_{\, 112} \, = \, \fr{1}{3}\, ( 2\, T_{\, 12} \, \L_1 \, + \, T_{\, 11} \, \L\, _2) \, ,      \\
&\d \, \cH_{\, 122} \,  =\, \fr{1}{3}\, (T_{\, 22} \, \L_1 \, + \, 2\, T_{\, 12} \, \L\, _2) \, ,     \\
&\d \, \cH_{\, 222} \, = \, T_{\, 22} \, \L_{\, 2} \, .
\end{split}
\ee
Out of these structures one could construct the scalar quantity (all indices contracted)
\be
{\cal{G}}_{\, 1111, 222} \, = \, T_{\, 11} \, T_{\, 22} \, \cH_{\, 112} \, - \, \fr{2}{3} \, T_{\, 12} \, T_{\, 22} \cH_{\, 111} \, - \, \fr{1}{3} \, T_{\, 11} \, T_{\, 11} \, \cH_{\, 222}\, ,
\ee
s.t. $\d \, {\cal{G}}_{\, 1111, 222} \equiv 0$. The existence of such a gauge-invariant combination implies the absence of enough gauge freedom allowing to remove all the fields $\cH_{\, ijk}$, if they were to be considered as independent. 

\end{appendix}


\end{document}